\documentclass[prc,twoside,superscriptaddress,showkeys,showpacs,twocolumn]{revtex4}%

\usepackage{graphicx, latexsym, amssymb, amsmath, color, multirow, mathrsfs, CJK, ifpdf}
\usepackage[section]{placeins}
\usepackage{amsmath}
\usepackage{amsfonts}
\usepackage{amssymb}
\usepackage{graphicx}
\newcommand{\beq}{\begin{equation}}
\newcommand{\eeq}{\end{equation}}
\newcommand{\beqn}{\begin{eqnarray}}
\newcommand{\eeqn}{\end{eqnarray}}
\newcommand{\bea}{\begin{array}}
\newcommand{\eea}{\end{array}}
\newcommand{\bsub}{\begin{subequations}}
\newcommand{\esub}{\end{subequations}}
\newcommand{\bpm}{\begin{pmatrix}}
\newcommand{\epm}{\end{pmatrix}}

\newcommand{\scr}[1]{{\mathscr #1}}
\newcommand{\ff}[1]{\frac{1}{#1}}

\newcommand{\lrb}[1]{\left(#1\right)}

\newcommand{\tabref}[1]{{Table \ref{#1}}}
\newcommand{\figref}[1]{{Fig. \ref{#1}}}

\newcommand{\sigl}{{\sigma\text{-}\Lambda}}
\newcommand{\omel}{{\omega\text{-}\Lambda}}

\begin{document}
\title{{Hyperon effects in covariant density functional theory with recent astrophysical observations}}

\author{Wen Hui Long}\email{longwh@lzu.edu.cn}
\affiliation{School of Nuclear Science and Technology, Lanzhou University, Lanzhou 730000, China}
\affiliation{Department of Physics, Tohoku University, Sendai 980-8578, Japan}
\author{Bao Yuan Sun}
\affiliation{School of Nuclear Science and Technology, Lanzhou University, Lanzhou 730000, China}
\author{Kouichi Hagino}
\affiliation{Department of Physics, Tohoku University, Sendai 980-8578, Japan}
\author{Hiroyuki Sagawa}
\affiliation{Center for Mathematical Sciences, University of Aizu, Aizu-Wakamatsu, Fukushima 965-8580, Japan}
\date{\today}

\begin{abstract}
Motivated by recent observational data, the equations of state with the inclusion of strangeness-bearing $\Lambda$-hyperons and the corresponding properties of neutron stars are studied, based on the covariant density functional (CDF) theory. To this end,
we specifically employ the density dependent relativistic Hartree-Fock (DDRHF) theory and the relativistic mean field theory (RMF). The inclusion of $\Lambda$-hyperons in neutron stars shows substantial effects in softening the equation of state. Because of the extra suppression effect originated from the Fock channel, large reductions on both the star mass and radius are predicted by the DDRHF calculations. It is also found that the mass-radius relations of neutron stars with $\Lambda$-hyperons determined by DDRHF with the PKA1 parameter set are in fairly good agreement with the observational data where a relatively small neutron stars radius is required. Therefore, it is expected that the exotic degrees of freedom such as the strangeness-bearing structure may appear and play significant roles inside the neutron stars, which is supported further by the systematical investigations on the consistency between the maximum
neutron star mass and $\Lambda$-coupling strength.
\end{abstract}

%
%
\maketitle


As the natural laboratories in the universe for nuclear and particle physics, neutron stars \cite{Lattimer:2004} have fascinated much effort concentrated on exploring the equation of state (EoS) of baryonic matter at low temperature and high density \cite{Glendenning:2000, Weber:2007PPNP}. Specifically the mass of the observed neutron stars brings a strong constraint on the behavior of EoS at supranuclear density. The most precise measurements for the neutron star mass are determined to be less than 1.5$M_\odot$ from the timing observations of radio binary pulsars \cite{Thorsett:1999APJ}, which has remained for many years to constraint the EoS. However, the existence of more massive compact stars is now unveiled by some evidence \cite{Lattimer:2007}. In a survey with the Arecibo telescope, an eccentric binary millisecond pulsar PSR J1903+0327 was found with an unusually high mass value $(1.74\pm0.04)M_\odot$ \cite{Champion:2008Science}. Recently a much larger pulsar mass of $(1.97\pm0.04)M_\odot$ was measured using Shapiro delay for the binary millisecond pulsar J1614-2230 \cite{Demorest:2011}. All of these new data imply a stiff EoS of strongly interacting matter at high densities, which need further check on the new developed land- and space-based observatories.

So far, there still exists considerable theoretical uncertainty on the EoS at supranuclear densities due to the poorly constrained many-body interaction, consequently deducing very different maximum mass and radius for a beta-stable neutron star. As indicated by prior study of neutron star based on the density dependent relativistic Hartree-Fock (DDRHF) theory \cite{BYSun:2008}, the maximum mass predicted by the covariant density functional (CDF) calculations \cite{ Walecka:1974, Long:2006} lies between 2$M_\odot$ and $2.8M_\odot$. The Corresponding EoSs deviate remarkably from each other at high density region. In the center of neutron stars, the density is generally considered as high as 5 to 10 times the nuclear equilibrium (saturation) density $\rho_0\thickapprox0.16~{\rm fm}^{-3}$ of neutrons and protons found in laboratory nuclei. At such high density, exotica such as strangeness-bearing baryons, condensed mesons, or even deconfined quarks could come into existence \cite{Lattimer:2004}, which may play significant roles in determining the EoS.

Besides the maximum mass limits, the mass-radius relation of neutron stars is also constrained by the recent observations, which leads to another strong restriction to the EoS. Recently, a relatively soft EoS and symmetry energy is predicted in the vicinity of the nuclear saturation density from careful analyses for three Type-I X-ray bursters with photospheric radius expansion and three transient low-mass X-ray binaries, which leads to a relatively small neutron star radius \cite{Ozel:2010PRD, Steiner:2010APJ}. Such observations are inconsistent with several commonly used equations of state that account only for nucleonic degrees of freedom. It is argued that they would be produced by including degrees of freedom beyond nucleons, e.g., hyperons, mesons, and quarks, or possibly produced by a better description of nucleonic interactions \cite{Ozel:2010PRD}. As a possible solution without exotic degrees of
freedom, the relativistic mean-field (RMF) models is recalibrated with a soft behavior of the symmetry energy around saturation density \cite{Fattoyev:2010PRC.025805, Fattoyev:2010PRC.055803}.

Compared to other CDF models, significant improvements have been obtained with DDRHF theory \cite{Long:2006, Long:2007} in describing the relativistic symmetry conservation \cite{Long:2006PS, Long:2007, Liang:2010}, the consistency of the isospin dependence in nuclear shell evolution \cite{Long:2008, Long:2009}, the exotic structures \cite{Long:2010a, Long:2010b}, and excited modes \cite{Liang:2008}. In the previous study \cite{BYSun:2008}, the significant contributions to the symmetry energy have been found from the Fock terms in the isoscalar $\sigma$ and $\omega$ couplings, and the neutron star properties determined by DDRHF theory are shown in fairly good agreement with the observational data.

In this paper, we continue the preivous work along the same line and study the roles of exotica in neutron stars based on the DDRHF theory. It is generally believed that {hyperons appear around twice the normal nuclear matter density in neutron star matter \cite{Shen:2002PRC.035802, Ishizuka:2008JPG, Sugahara:1994PTP}}. The first hyperon to appear should be $\Lambda$ as it is the lightest {one} with an attractive potential in nuclear matter. From the experimental binding energies of single-$\Lambda$ hypernuclei, the potential depth of $\Lambda$ in nuclear matter is estimated to be $\thicksim30$ MeV \cite{Shen:2006PTP}. Motivated by the recent astrophysical observation, as a preliminary attempt it is thus interesting to introduce the strangeness degree of freedom associated with $\Lambda$-hyperon into the DDRHF theory and to investigate the corresponding neutron star properties.


Within DDRHF, the baryons are described as point-like particles and interact with each other by exchanging mesons (the isoscalar $\sigma$ and $\omega$ as well as isovector $\rho$ and $\pi$) and photons ($A$). The $\Lambda$-hyperon ($\Lambda = uds$), whose strangeness $S = -1$, isospin $I$ = 0 and spin-parity $J^P=\ff2^+$, participates only in the interactions propagated by the isoscalars, i.e., the $\sigma$- and $\omega$-mesons. These interactions can be described by the following Lagrangian density,
\begin{equation}\label{Lagrangian_L}
\scr L_\Lambda = \bar \psi_\Lambda\lrb{i\gamma^\mu \partial_\mu - M_\Lambda - g_\sigl \sigma - g_\omel\gamma^\mu\omega_\mu}\psi_\Lambda,
\end{equation}
where $M_\Lambda$ denotes the mass of $\Lambda$-hyperon ($\psi_\Lambda$). From the Lagrangian density (\ref{Lagrangian_L}) the contributions to the energy functional, as well as the Dirac equation for $\Lambda$-hyperon, can be determined similarly as nucleon  \cite{Bouyssy:1987, Long:2006, Long:2007,BYSun:2008}.

For the beta-stable stellar matter containing nucleons, $\Lambda$-hyperons, electrons, and muons, the chemical equilibrium conditions require that
\begin{align}
\mu_p = &\mu_n - \mu_e, \\
\mu_\Lambda=&\mu_n, \label{chemical}\\
\mu_\mu = &\mu_e
\end{align}
where the chemical potential $\mu_n$, $\mu_p$, $\mu_\Lambda$, $\mu_\mu$, and $\mu_e$ are determined by the relativistic energy-momentum relation at Fermi momentum $p = k_F$ \cite{BYSun:2008},
\begin{align}
\mu_i = & \Sigma_0(k_{F,i}) + E^*(k_{F,i}), &\mu_\lambda = & \sqrt{k_{F,\lambda}^2 + m_\lambda^2}.
\end{align}
In above expressions, $i$ denotes the baryons, i.e., neutron ($n$), proton ($p$), and $\Lambda$-hyperon, and $\lambda$ represents the leptons, electron ($e^-$) and muon ($\mu^-$). Further combined with the baryon density conservation and charge neutrality, i.e.,
\begin{align}
\rho_b = & \rho_n + \rho_p + \rho_\Lambda, &
\rho_p = & \rho_\mu + \rho_e,
\end{align}
the ratios of baryons and leptons can be obtained from the self-consistent calculations for the stellar matter with given baryon density $\rho_b$. The equation of state (EoS) -- the relation between the pressure and energy density for stellar matter, is then determined by DDRHF as well as other CDF calculations, in which the baryons interact by exchanging mesons whereas the leptons are described as free fermions.

With the EoS of the stellar matter, the structure of a static, spherically symmetric, and relativistic star can be determined by solving the Tolman-Oppenheimer-Volkov (TOV) equations \cite{Oppenheimer:1939, Tolman:1939}.Similarly as in Ref. \cite{BYSun:2008}, the EoS in the low density region ($\rho_b<0.08$fm$^{-3}$) will be provided by the BPS \cite{Baym:1971AJ} and BBP \cite{Baym:1971NPA} models. For a given central density $\rho(0)$ or central pressure $P(0)$, the input EoS leads to the unique solution of the TOV equations.


In the following the theoretical calculations are systematically performed on the platform of CDF theory \cite{Serot:1986, Bouyssy:1987, Meng:2006, Long:2006, Long:2007}, specifically the DDRHF theory with effective interactions PKA1 \cite{Long:2007} and PKO3 \cite{Long:2008}, and the RMF theory with NL-SH \cite{Sharma:1993}, PK1 \cite{Long04}, TW99 \cite{Typel:1999}, and PKDD \cite{Long04}. {In \tabref{tab:details} are summarized the details of the selected effective Lagrangians.} For the RMF calculations, only the Hartree contributions are involved in the baryon-baryon interactions, whereas both Hartree and Fock terms are taken into account by DDRHF. In NL-SH and PK1, the non-linear self-couplings of $\sigma$- and $\omega$-mesons \cite{Sugahara:1994, Boguta:1977, Long04} are introduced to evaluate the in-medium effects of nuclear interaction. In TW99, PKDD, PKA1, and PKO3, such effects are {introduced} by the density dependence in meson-nucleon couplings \cite{Lenske:1995, Brockmann:1992} and the same density dependent behaviors are utilized in corresponding meson-$\Lambda$ coupling channels, {i.e., the meson-baryon coupling constants are treated as functions of baryon density ($\rho_b = \rho_n +\rho_p +\rho_\Lambda$)}. {As proved by prior studies \cite{BYSun:2008}, such prescriptions on the in-medium effects are still reasonable for exploring the EoS at supranuclear density as well as in describing the neutron stars.}

\begin{table}[htpb]
\setlength{\tabcolsep}{4pt}
\caption{Details of the selected CDF effective Lagrangians. }\label{tab:details}
\begin{tabular}{c|c|cc|c}\hline\hline
 & Fock& non-linear & non-linear & density dependent \\
 & term& $\sigma$ term & $\omega$ term & couplings \\
\hline
PKA1 & yes & no & no & yes \\
PKO3 & yes & no & no & yes \\
\hline
NLSH & no & yes & no & no \\
PK1 & no & yes & yes & no \\
TW99 & no & no & no & yes \\
PKDD & no & no & no & yes \\
\hline\hline
 \end{tabular}
\end{table}

In the CDF calculations if not specified the proportions between meson-$\Lambda$ and meson-nucleon couplings are fixed as $g_\sigl/g_\sigma = 0.600$ and $g_\omel/g_\omega = 0.653$ \cite{Glendenning:1993}, and the masses of $\Lambda$-hyperon, electron, and muon are chosen as $M_\Lambda=1115.0$ MeV, $m_e = 0.511$ MeV, and $m_\mu = 105.7$ MeV. In the Lagrangian density (\ref{Lagrangian_L}) we neglect the strangeness concerned baryon-baryon interactions, which still remain to be an open question. In fact, several recent observations of double-$\Lambda$ hypernuclei indicate that the effective $\Lambda\Lambda$ interaction should be considerably weaker than earlier evaluations \cite{Nakazawa:2010NPA.207}.

\begin{figure}
\ifpdf
\includegraphics[width=0.48\textwidth]{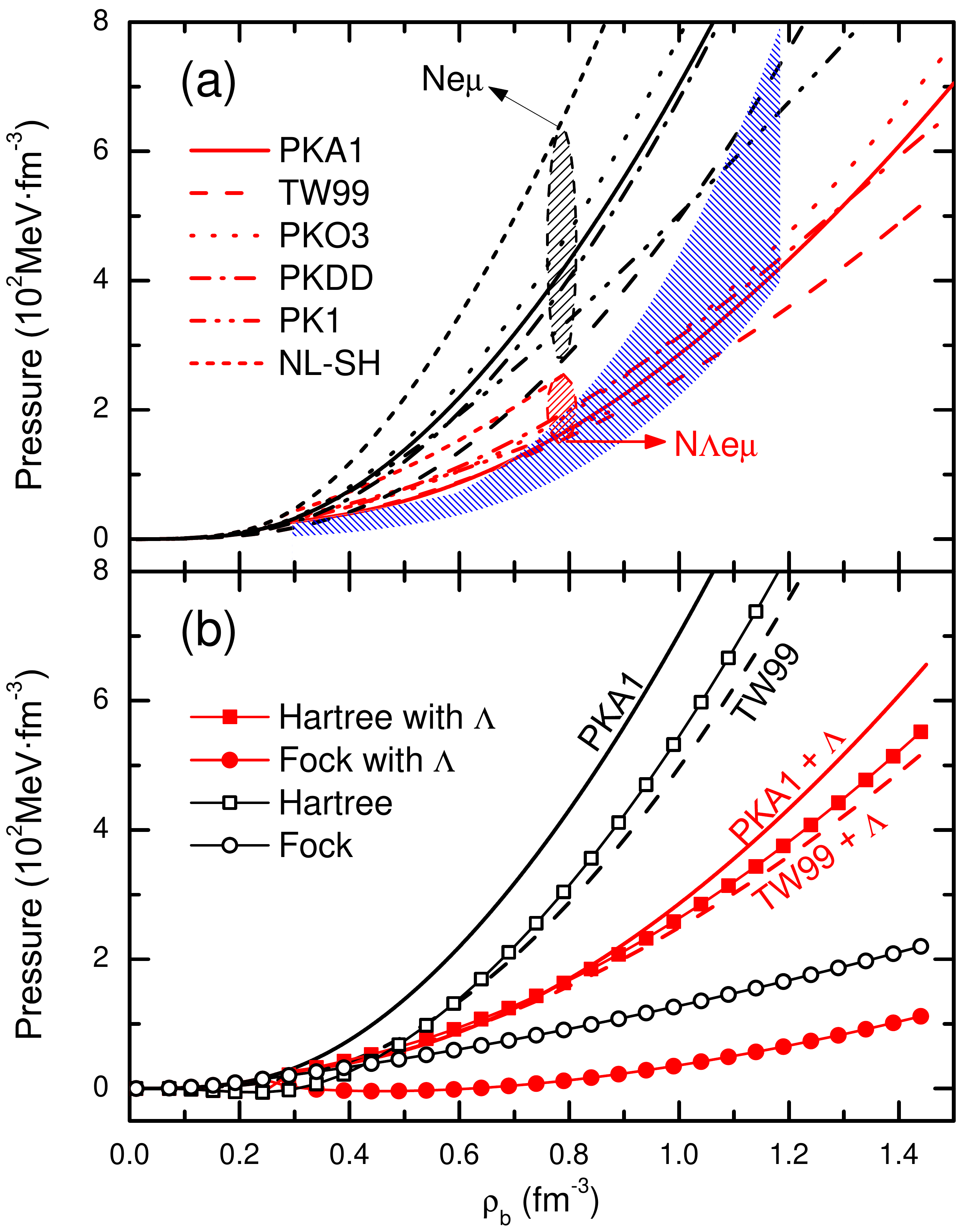}
\else
\includegraphics[width=0.48\textwidth]{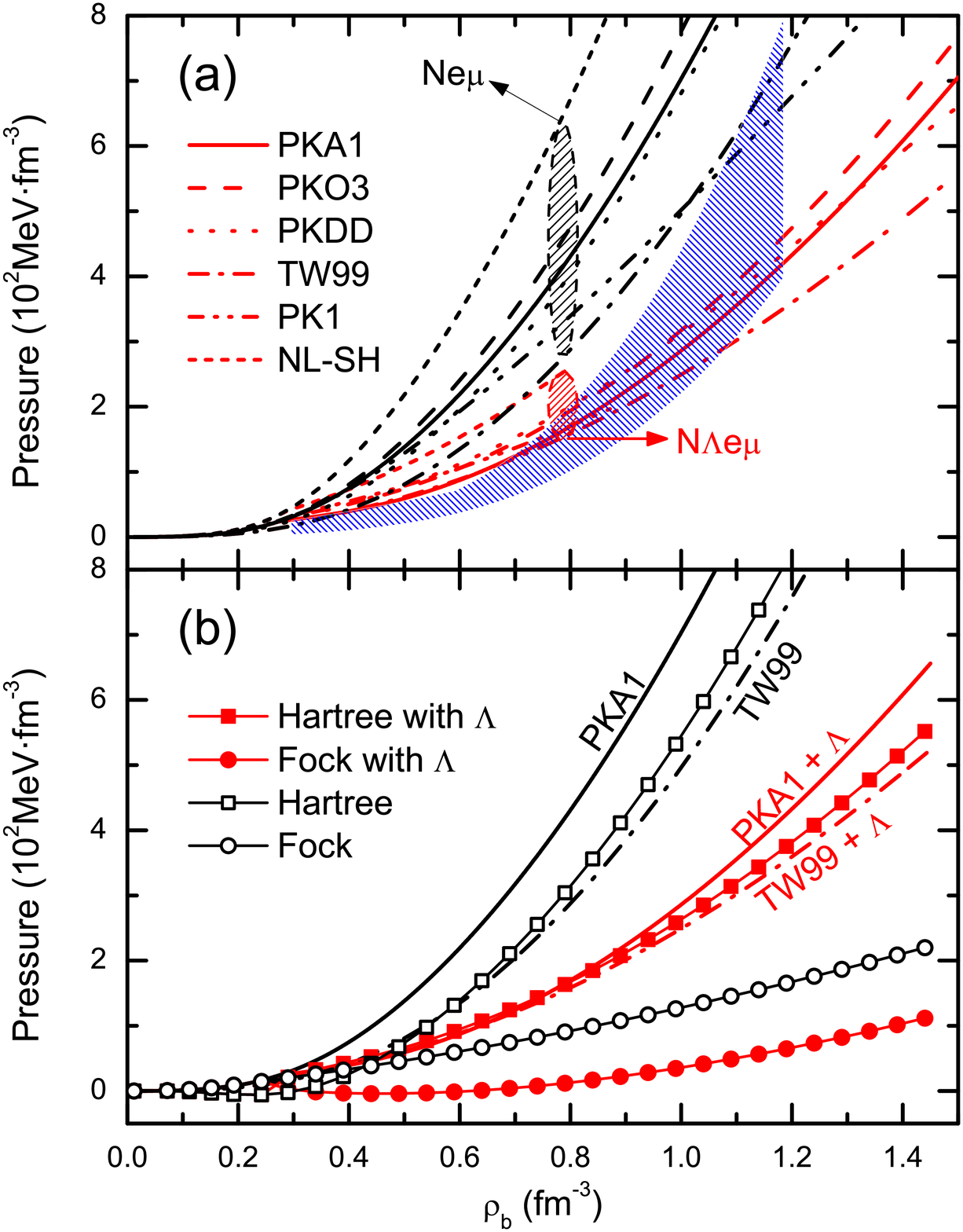}
\fi
\caption{(Color online) The pressure as a function of the baryonic density $\rho_b$ (fm$^{-3}$) with different CDF effective interactions for the stellar matter containing nucleons, $\Lambda$ hyperons, electrons and muons (red curves), as compared to the one without $\Lambda$ hyperons (black curves). The corresponding contribution from Hartree and Fock channel with PKA1 are shown in the lower panel in comparison with TW99. See the text for details.}\label{fig:EoS}
\end{figure}

With the selected CDF effective Lagrangians, we first study the equation of state for the beta-stable stellar matter containing nucleons, $\Lambda$-hyperons, electrons and muons (or $N\Lambda e\mu$ for short), in comparison with those without $\Lambda$-hyperons (or $Ne\mu$ for short). \figref{fig:EoS}(a) shows the pressures of neutron star matter as functions of the baryon density $\rho_b$ with {the selected} CDF effective interactions. For the $Ne\mu$ matter, there exist substantial deviations (black shadowed ellipse) {between the EoSs}, among which NL-SH presents the hardest EoS and TW99 gives the softest one. When the $\Lambda$-hyperon degree of freedom is introduced, the EoSs become much softer above the critical density ($\sim0.3$ fm$^{-3}$), where $\Lambda$-hyperons start to appear in stellar matter. In addition, the deviations (red shadowed ellipse) {on the} EoSs are considerably reduced as well. The distinctly stiff behavior of EoS with NL-SH effective interaction is mainly due to the absence of the non-linear $\omega$ self-couplings \cite{Sugahara:1994}. It is also found that PKA1 presents rather harder EoS than TW99 in the $Ne\mu$ matter, while with $\Lambda$-hyperons PKA1 and TW99 tend to provide almost identical EoS until {$\rho_b\thickapprox 0.8$ fm$^{-3}$,} where the deviation emerges again.

In fact, such features on the EoS can be understood qualitatively from the chemical equilibrium between nucleons and hyperons [see Eq.(\ref{chemical})]. {For harder EoS, the neutron chemical potential goes up faster with the density increasing, such that more neutrons will be transferred into $\Lambda$-hyperons}. {Because the $\omega$-meson couplings with $\Lambda$-hyperons, which play dominant role at high density, are generally weaker than those with nucleons as repulsions}, the appearance of $\Lambda$-hyperon will soften the EoS evidently. Thus, {more stiffer a EoS of the $Ne\mu$ matter is, more} effort to soften the EoS is made by $\Lambda$-hyperon. When the density reaches at rather large values, the ratios {of} $\Lambda$-hyperons and nucleons become stable so that the deviations among the effective Lagrangians appear again. Even though, such deviations have been diminished evidently due to the fairly large proportion of $\Lambda$-hyperon at high density. Compared to the constraint [shadowed area in \figref{fig:EoS}(a)] suggested in Ref. \cite{Ozel:2010PRD} for cold matter, the EoSs of $N\Lambda e\mu$ matter {show} much better consistency with the constraint than those for $Ne\mu$ matter.

To illustrate the influence of the Fock term on the EoS for the $Ne\mu$ and $N\Lambda e\mu$ matter, the pressures of neutron star matter with DDRHF effective interaction PKA1 and DDRMF one TW99 are replotted in \figref{fig:EoS}(b) as functions of the baryon density. {The contributions from the Hartree and Fock channels are shown as well}. Within DDRHF, one can find that the {Hartree terms provide} the dominant contributions to the pressures in comparison with {the Fock terms} in both $Ne\mu$ and $N\Lambda e\mu$ matter. It is also seen that {the Hartree contributions in PKA1 are nearly identical with TW99 and} the difference in the EoSs comes significantly from the Fock term in PKA1. {Compared to the $Ne\mu$ matter, the occurrence of $\Lambda$-hyperons remarkably suppresses the Fock contributions in the $N\Lambda e\mu$ matter. This is mainly due to the effects of $\omega$ couplings in the Fock channel. In $Ne\mu$ matter the Fock terms in the $N$-$\omega$ couplings contribute as fairly strong repulsion at high density. {Whereas} in $N\Lambda e\mu$ matter the $\Lambda$-$\omega$ couplings in Fock channel represent as attraction, which substantially softens the EoS.} {As a result} nearly identical behavior of EoSs is {provided by} PKA1 and TW99 {in the density region about $2\sim 5\rho_0$}, {showing better consistency with the constraint \cite{Ozel:2010PRD} than other CDF effective Lagrangians. }

\begin{figure}
\ifpdf
\includegraphics[width=0.48\textwidth]{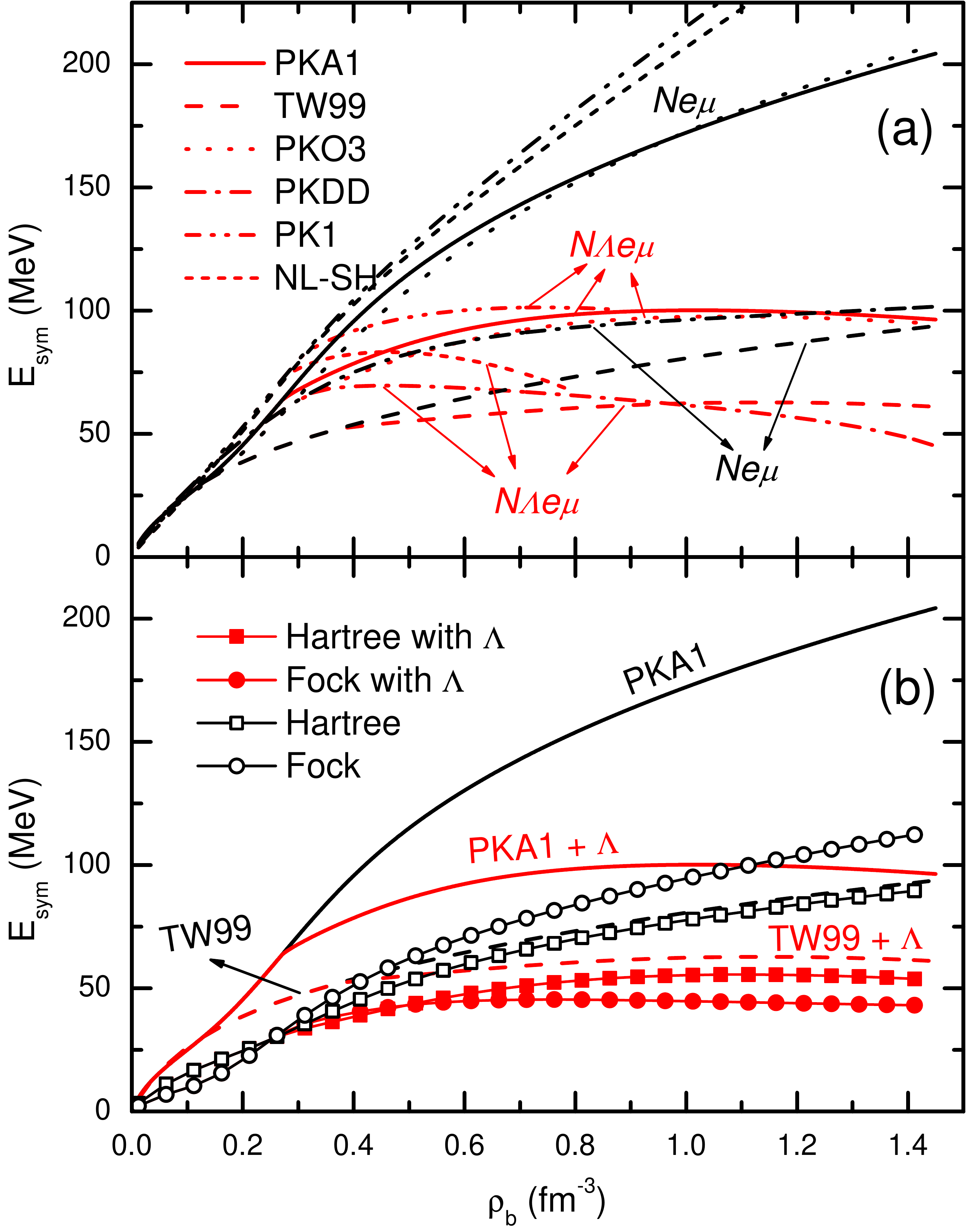}
\else
\includegraphics[width=0.48\textwidth]{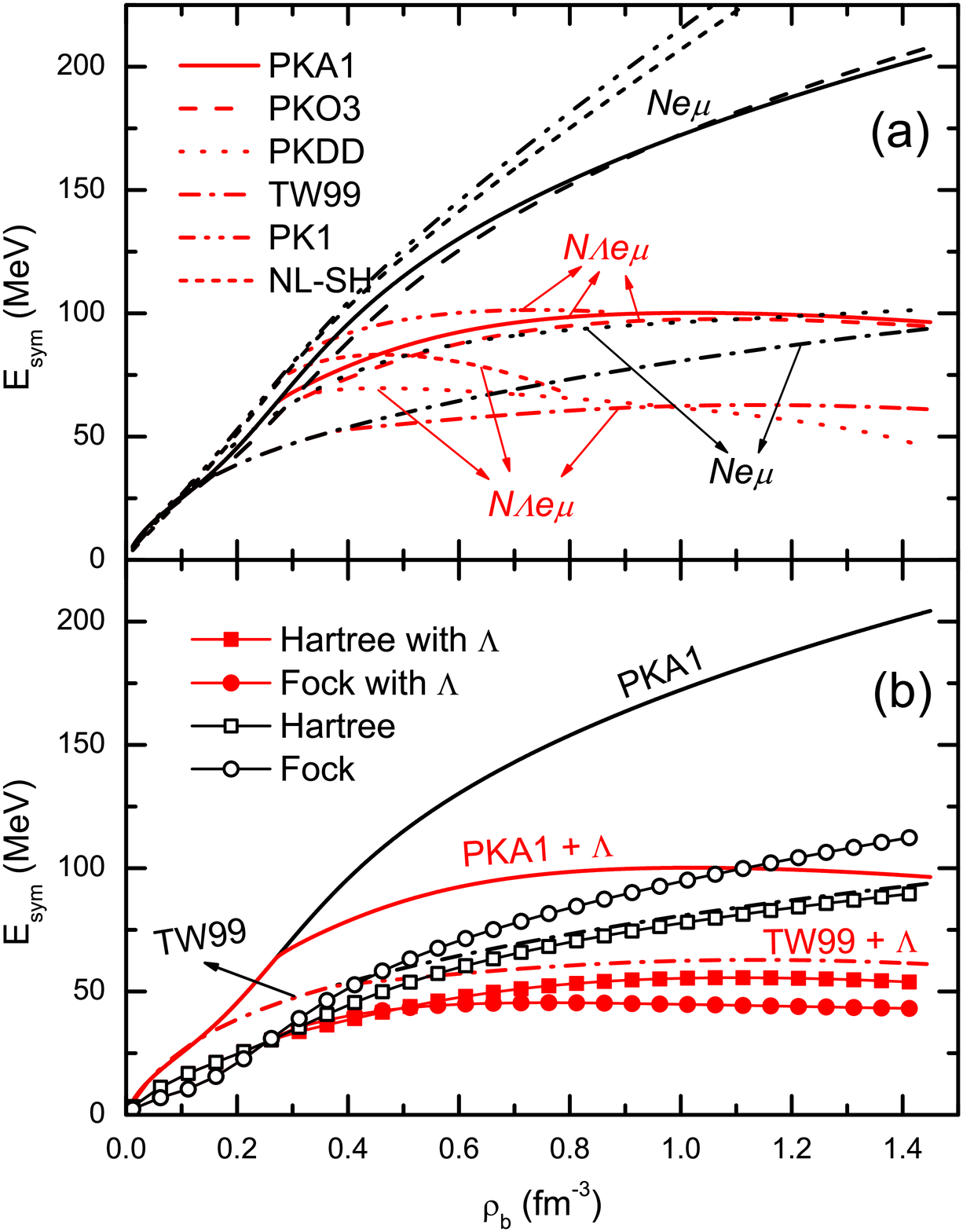}
\fi
\caption{(Color online) The symmetry energy $E_{sym}$ (MeV) as a function of the baryonic density $\rho_b$ (fm$^{-3}$) with different CDF effective interactions for the stellar matter containing nucleons, $\Lambda$ hyperons, electrons and muons (red curves), as compared to the one without $\Lambda$ hyperons (black curves). The corresponding contribution from Hartree and Fock channel with PKA1 are shown in the lower panel in comparison with TW99. See the text for details.}\label{fig:Sym}
\end{figure}

The symmetry energy is an important quantity for illustrating the property of asymmetric nuclear matter. In general, the energy per particle of asymmetric nuclear matter, $E(\rho_b,\beta)$, can be expanded in {the} Taylor series {with respect to the asymmetry parameter $\beta=(\rho_n-\rho_p)/(\rho_n+\rho_p)$},
\begin{equation}\label{EsParabolicLaw}
E(\rho_b,\beta)=E_0(\rho_b)+\beta^2E_{sym}(\rho_b)+\cdots.
\end{equation}
The function $E_0(\rho_b)$ is the binding energy per particle in symmetric nuclear matter. The empirical parabolic law in Eq.~(\ref{EsParabolicLaw}) is confirmed to be reasonable throughout the range of the asymmetry parameter values especially at low density. {As a reasonable approximation}, one can extract the symmetry energy $E_{sym}(\rho_b)$ for the beta-stable stellar matter by
\begin{equation}
E_{sym}(\rho_b) = \frac{E(\rho_b,\beta)-E_0(\rho_b)}{\beta^2}.
\end{equation}

\figref{fig:Sym}(a) shows the symmetry energy of neutron star matter as functions of the baryon density $\rho_b$ with different CDF effective interactions. For the $Ne\mu$ matter, sizable enhancements of $E_{sym}$ are seen in the high density region with respect to $E_{sym}$ at the saturation denstiy. When the $\Lambda$-hyperon is introduced, the {symmetry energies} become much softer{, about 50\% reduced at high densities}. Similar as the results in \figref{fig:EoS}, the deviations of $E_{sym}$ among different EoSs are also reduced in the $N\Lambda e\mu$ matter. For the $Ne\mu$ matter, the Fock terms of $\sigma$- and $\omega$-couplings exhibit significant contributions to the symmetry energy, which leads to a stronger density dependence {in DDRHF than in DDRMF at high density} \cite{BYSun:2008}. {While} in the $N\Lambda e\mu$ matter, both Hartree and Fock contributions (e.g., in PKA1) are much reduced by $\Lambda$-hyperons. In \figref{fig:Sym}(b) one may find that the Hartree contributions in PKA1 are essentially identical with TW99 and correspondingly the Fock terms provide about 2-3 times reduction as the Hartree ones with the occurrence of $\Lambda$-hyperons, leading to notably soft symmetry energy at high densities.

\begin{figure}
\ifpdf
\includegraphics[width=0.48\textwidth]{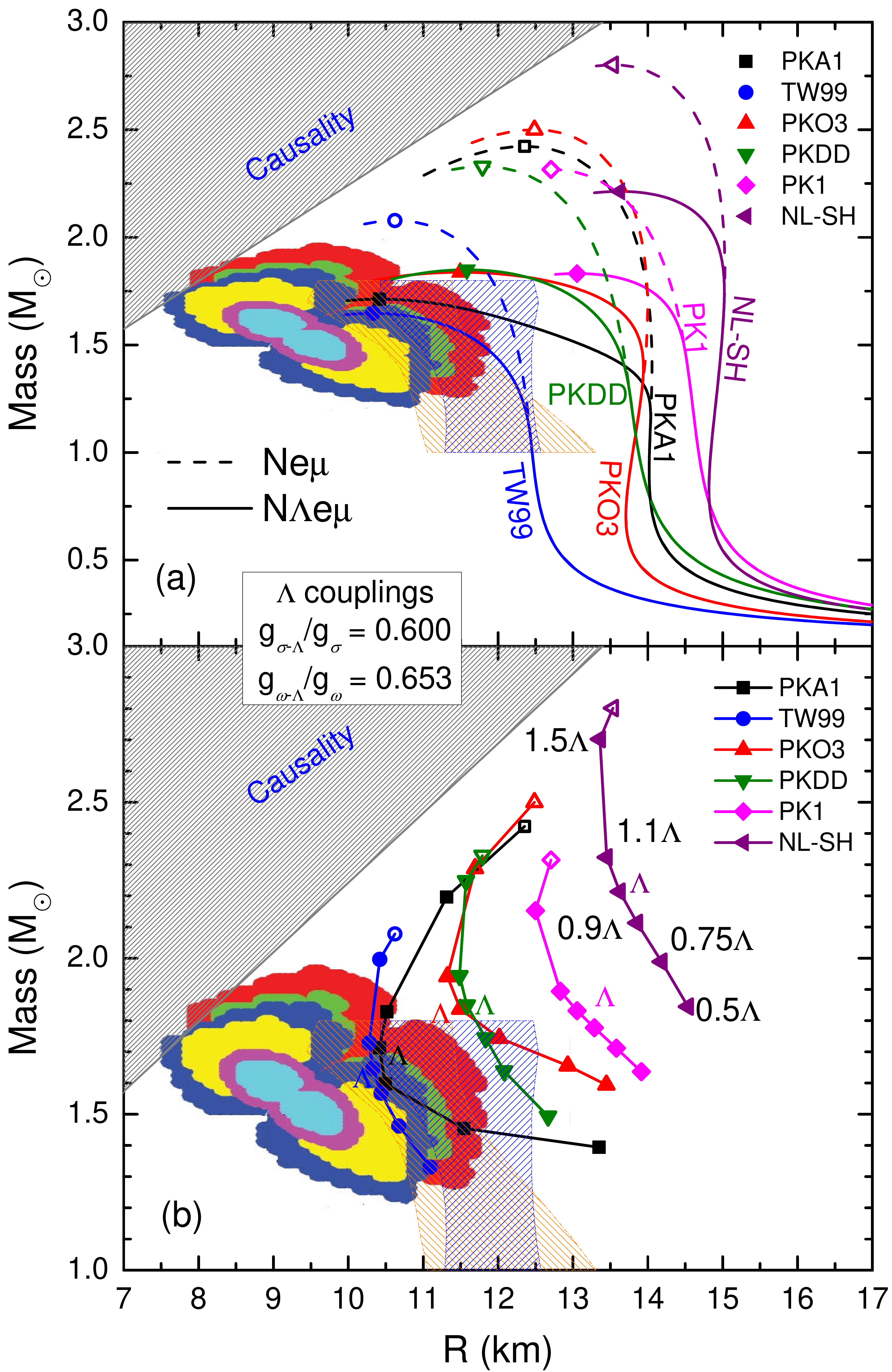}
\else
\includegraphics[width=0.48\textwidth]{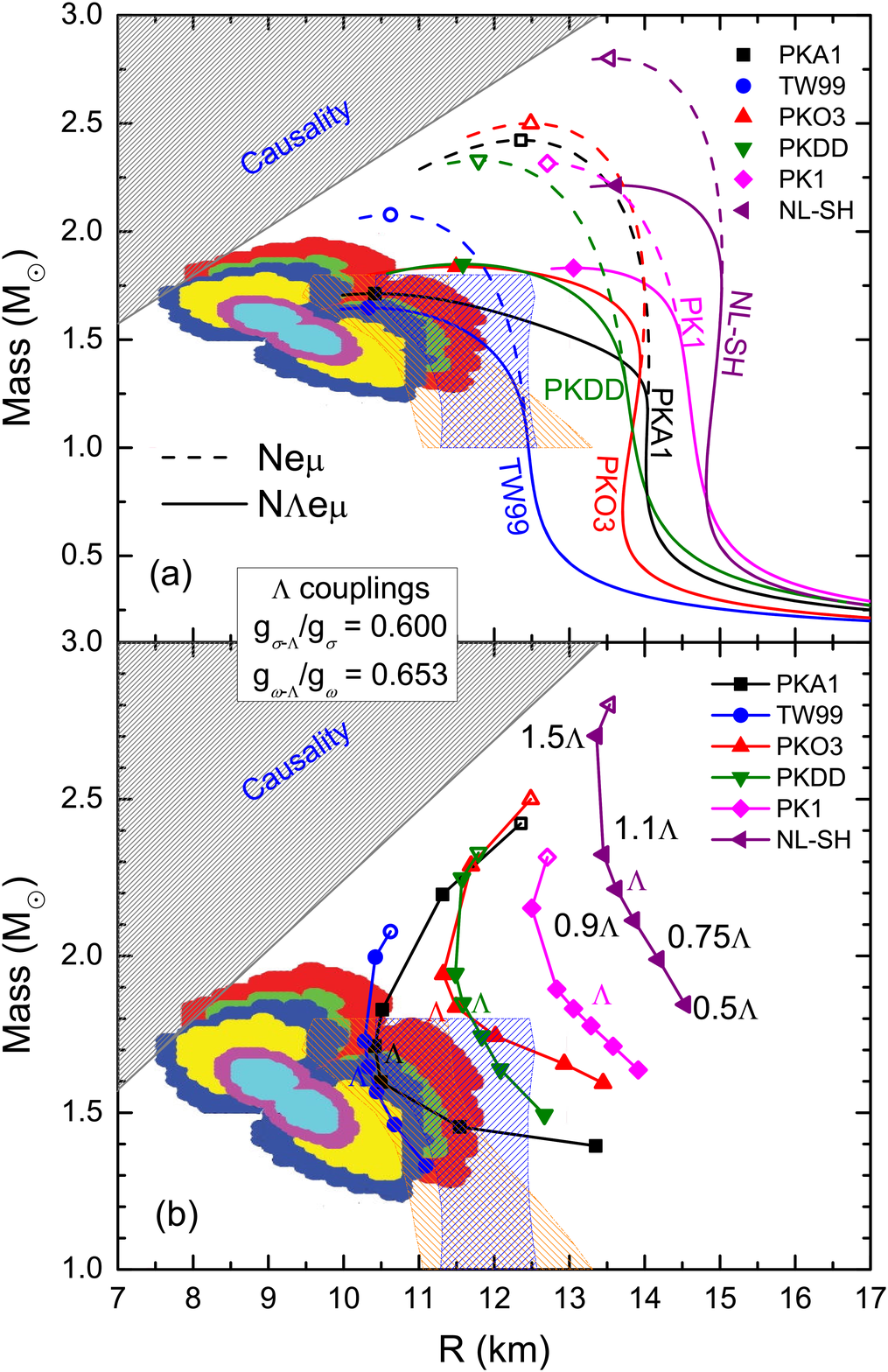}
\fi
\caption{(Color online) The mass-radius relations for the neutron stars with $\Lambda$ (solid lines) and without $\Lambda$ (dashed lines) hyperons. The symbols denote the neutron stars with maximum masses. Region excluded by causality is indicated with gray color. For comparison the observational constraints from three neutron stars in the binaries 4U 1608$-$248 (green/red), EXO 1745$-$248 (yellow/blue), and 4U 1820$-$30 (cyan/magenta) in Ref. \cite{Ozel:2010PRD} are shown as the 1- and 2-$\sigma$ confidence contours, and the later analyses in Ref. \cite{Steiner:2010APJ} are denoted by the shadowed areas with two different models of photospheric radius. In the lower panel are shown the mass \& radius for the neutron stars with maximum masses determined with different $\Lambda$ hyperon coupling strength (filled symbols), as compared to the ones without $\Lambda$ hyperons (open symbols). See the text for details.} \label{fig:TOV}
\end{figure}

Taking the EoSs in \figref{fig:EoS}(a) as the input, the systematic properties of neutron stars such as the mass-radius relation can be obtained by solving the TOV equations. In \figref{fig:TOV}(a) are shown the mass-radius relations for the neutron stars which consist of the $N\Lambda e\mu$ matter (solid lines), compared to the $Ne\mu$ matter (dashed lines). Consistently with the EoSs in \figref{fig:EoS}, the inclusion of $\Lambda$-hyperon brings substantial reduction on the neutron star mass. When $\Lambda$-hyperons appear in neutron star, the mass-radius relations deviate from those without $\Lambda$-hyperons and bend down to smaller masses and radii. {Close to the maximum points (in symbols) the mass-radius relations tend stable, which is also different from the calculations without $\Lambda$-hyperon. This can be interpreted by the behaviors of symmetry energy at high density. In $Ne\mu$ matter the symmetry energies increase with the density almost linearly. While with the inclusion of $\Lambda$-hyperons the symmetry energies bend down to be stable or even slightly decreasing when the density goes high, in a consistency with the mass-radius relations [see \figref{fig:TOV}(a)].}

{Moreover, the CDF calculations for the $Ne\mu$ matter predict quite different mass for neutron stars. However, when $\Lambda$-hyperon is included, the deviations of the maximum masses among different EoSs are considerably reduced, namely, the CDF effective Lagrangians except NL-SH predict vicinal maximum mass. Comparing PKO3 with PKDD and PK1, one can find that nearly identical maximum masses are provided by PKDD and PK1 in both cases whereas PKO3 gives larger value in the case of the $Ne\mu$ matter. As mentioned above this can be understood well from the effects of the Fock terms in softening the EoS and thus reducing the maximum mass of neutron stars.}

Extracted from \figref{fig:TOV}(a), \tabref{tab:bulksM} shows the mass ($M_\odot$), radius (km), and central density $\rho_c$ (fm$^{-3}$) for neutron stars with the maximum mass, as well as the threshold densities $\rho_b^\Lambda$ (fm$^{-3}$) and $\rho_b^\mu$ (fm$^{-3}$) for $\Lambda$ and muon emergence. For comparison, the results given by the calculations without $\Lambda$-hyperon are shown in the lower panel. From \tabref{tab:bulksM} one can clearly see the mass reduction induced by the occurrence of $\Lambda$-hyperons, especially in the CDF calculations with the Fock terms, i.e., the DDRHF calculations (PKA1 and PKO3), which present the mass reduction about 0.7$M_\odot$. In the RMF calculations the mass reductions range from 0.4 to 0.5 $M_\odot$, except NL-SH which gives the reduction about 0.6$M_\odot$.

\begin{table}
\setlength{\tabcolsep}{4pt}
\caption{Maximum mass $M_{\max}$ ($M_\odot$), corresponding radii $R_{\max}$ (km) and central density $\rho_c$ (fm$^{-3}$) for neutron stars, as well as the threshold densities $\rho_b^\Lambda$ (fm$^{-3}$) and $\rho_b^\mu$ (fm$^{-3}$) of $\Lambda$ and muon. For comparison, the quantities for neutron stars without $\Lambda$ hyperons are shown in the lower panel. The results are calculated by DDRHF with PKA1 and PKO3, and RMF with PKDD, TW99, PK1, and NL-SH. } \label{tab:bulksM}
 \begin{tabular}{c|rr|rr|rr}\hline\hline
          & PKA1 & PKO3 & PKDD & TW99 &  PK1 & NL-SH \\ \hline
$M_{\max}$&1.713 & 1.837& 1.849& 1.647& 1.832& 2.213 \\
$R_{\max}$&10.425&11.495&11.583&10.333&13.048&13.633 \\
$\rho_c$  &1.314 & 1.048& 1.024& 1.292& 0.786& 0.700 \\ \hline
$\rho_b^\Lambda$&0.272&0.284&0.322&0.368&0.306&0.282 \\
$\rho_b^\mu$&0.118&0.122&0.108&0.116&0.110&0.114\\ \hline\hline
$M_{\max}$& 2.423& 2.500& 2.329& 2.078& 2.315& 2.802\\
$R_{\max}$&12.354&12.487&11.798&10.632&12.705&13.534\\
$\rho_c$  & 0.810& 0.780& 0.888& 1.104& 0.796& 0.650\\ \hline\hline
 \end{tabular}
\end{table}

The inclusion of $\Lambda$-hyperon also brings distinct effects on central density $\rho_c$. Among the selected effective Lagrangians PKA1 and TW99 predict the largest values of $\rho_c$, about 8 times of the saturation density ($\sim$ 0.16 fm$^{-3}$). Compared to those excluding $\Lambda$-hyperon, the central densities increase roughly {25\%$\thicksim$38\% (about $0.25\thicksim0.50$fm$^{-3}$) in the DDRHF calculations (PKO3 and PKA1)}, and {about 10\%$\thicksim$15\%} in the density dependent RMF calculations (PKDD and TW99). While the central density predicted by PK1 decreases with the $\Lambda$-hyperon inclusion and NL-SH
simply gives tiny changes in $\rho_c$.

For the radii of neutron stars with the maximum mass, distinct reductions are also found in the DDRHF calculations including $\Lambda$-hyperon, e.g., PKA1 predicts the star radius about 10.4 km, about 2~km smaller than the one without $\Lambda$-hyperon, and PKO3 provides {the reduction about 1~km}. In the RMF calculations, the neutron star radii are changed only slightly ($<0.3$km) with the inclusion of $\Lambda$-hyperon. Such discrepant behavior between DDRHF and RMF originates from the correlations between the radius of neutron stars and the symmetry energy. {In \tabref{tab:bulks} are shown the bulk quantities of symmetric nuclear matter at saturation density for the selected effective Lagrangians. As seen from the slope $L$ and curvature $K_{\text{sym.}}$ on the symmetry energy \cite{CheLW:2005PRC.064309}, the DDRMF calculations (PKDD and TW99) present rather soft symmetry energy and the slopes tend to decrease as density goes high due to the negative curvature $K_{\text{sym.}}$. In an accordance, small neutron star radii are obtained in the case of $N e\mu$ matter, especially for TW99. Similar consistence can also be seen from the calculations of DDRHF and NLRMF. In \figref{fig:Sym} it is clearly shown that the occurrence of $\Lambda$-hyperons brings substantial effects in reducing the slope of the symmetry energy, mainly from the Fock channel in DDRHF. Due to such extra suppression, consistently the reductions of the star radius are more dramatic than the case of RMF, as proved by the results in \tabref{tab:bulksM}. On the other hand it is also well demonstrated that the properties of neutron star are strongly correlated with the values of the symmetry energy. As seeing from the results in \figref{fig:Sym}(a), one should notice that the symmetry energies are dramatically changed with the occurrence of $\Lambda$-hyperon at high density so that the properties of symmetry energy at normal density are not enough to describe properly the size and mass of neutron stars.}

\begin{table}[htbp]\setlength{\tabcolsep}{0.5em}
\caption{Bulk properties of symmetric nuclear matter at saturation point, i.e., the saturation density $\rho_0$ (fm$^{-3}$), binding energy per particle $E/A$ (MeV), incompressibility $K$ (MeV), and symmetry energy $J$ (MeV) with its slope $L$ (MeV) and curvature $K_{\text{sym.}}$ (MeV). The results are provided by the CDF effective Lagrangians PKA1, PKO3, PKDD, TW99, NL-SH, and PK1.}\label{tab:bulks}
\begin{tabular}{c|rrr|rrr}\hline\hline
       & $\rho_0$~~~ & $E/A$~~&   K~~~ & $J$~~~&  $L$~~~&$K_{\text{sym.}}$ \\  \hline
 PKA1  &  0.160      & -15.83 & 229.96 & 36.02 & 103.50 & 213.23 \\
 PKO3  &  0.153      & -16.04 & 262.47 & 32.99 &  83.00 & 116.56\\  \hline
 PKDD  &  0.150      & -16.27 & 262.18 & 36.79 &  90.21 & -80.74\\
 TW99  &  0.153      & -16.25 & 240.27 & 32.77 &  55.31 &-124.68\\  \hline
 NL-SH &  0.146      & -16.35 & 355.43 & 36.12 & 113.66 &  79.72\\
 PK1   &  0.148      & -16.27 & 282.69 & 37.64 & 115.88 &  55.33\\
\hline\hline
\end{tabular}
\end{table}

In \figref{fig:TOV}, the 1- and 2-$\sigma$ confidence contours for the masses and radii of three neutron stars in the binaries 4U 1608$-$248 (green/red), EXO 1745$-$248 (yellow/blue), and 4U 1820$-$30 (cyan/magenta) are shown \cite{Ozel:2010PRD}, and a re-analyzed version recently done by another group is denoted by the shadowed areas with two different models of photospheric radius as well \cite{Steiner:2010APJ}. Different from other constraints like shown in Ref. \cite{BYSun:2008}, these new observations put a serious challenge to our understanding of neutron stars, where a relatively small values about 8.7-12.5~km is required for a $1.4M_\odot$ star, even smaller than a recent conclusion of 9.7-13.9~km from microscopic calculations based on chiral effective field theory interactions \cite{Hebeler:2010PRL.161102}. Therefore a very soft symmetry energy and EoS near and above several times of the saturation density are needed. In comparison with our calculations, it is found that the mass-radius relations for the neutron stars with $\Lambda$-hyperons given by PKA1 and TW99 are in perfect accordance with the observations, especially for the cases around the maximum mass. Because of a little {harder} EoSs at the density region about  $2\thicksim5\rho_0$, PKO3 and PKDD with $\Lambda$-hyperons just marginally cover the constraints, while the nonlinear RMF versions PK1 and NL-SH could not fulfill the constraints at all. In the cases without $\Lambda$-hyperons, all the curves are far away from the constraints. Hence, it is strongly suggested the exotic degrees of freedom such as the strangeness-bearing structure may appear inside the neutron stars. It is expected {and also found in the CDF calculations} that $\Lambda$-hyperon is the dominant constitution in the core of neutron star whereas neutron is strongly compressed to be less than 20\%, from which the role of the hyperon degree of freedom is well demonstrated in neutron stars.

In \figref{fig:EoS} and \figref{fig:TOV}(a) it is already shown that the $\Lambda$-hyperon plays an important role in softening the EoS and reducing the neutron star masses and radii {and leads to fairly good agreements with the constraints \cite{Ozel:2010PRD, Steiner:2010APJ}}. {One may notice that in the case of $N\Lambda e\mu$ matter none of the curves except NL-SH go through 2.0$M_\odot$, which is constrained by another observation \cite{Demorest:2011}. On the other hand if comparing the results provided by the selected effective Lagrangians it is also found that for the CDF calculations with Fock terms, i.e., the DDRHF ones, the inclusion of $\Lambda$-hyperon brings more distinct effects in reducing the neutron star masses.}

In all the above calculations the coupling strengths of $\Lambda$-hyperon are fixed to $g_\sigl/g_\sigma = 0.600$ and $g_\omel/g_\omega = 0.653$. For the $\Lambda$-hyperon, it only participates in the interaction mediated by the exchange of the isoscalar $\sigma$- and $\omega$-mesons, which respectively represent as strong attraction (repulsion) and repulsion (attraction) in Hartree (Fock) channels. It is well known that at high density the contributions from $\omega$-meson play the dominant role in determining the EoS, as well as the mass-radius relation for neutron stars. In DDRHF there exist substantial cancelations between the Hartree and Fock terms of the {$\omega$-$\Lambda$} coupling and such cancelations are somewhat equivalent to weakening the coupling strength. More distinct effects are therefore found in the DDRHF calculations with the inclusion of $\Lambda$-hyperon in softening the EoS, as well as reducing the neutron star mass.

Here we only consider the strangeness system degree of freedom associated with $\Lambda$-hyperon. Approximately and qualitatively, effectively other strangeness related contributions such as $\Sigma$- and $\Xi$-hyperons can be taken into account by modifying the $\Lambda$-coupling strength. In \figref{fig:TOV}(b) are shown the neutron stars with the maximum mass extracted from the CDF calculation with different $\Lambda$-coupling strengths, where the Greek symbol $\Lambda$ denotes the coupling strengths $g_\sigl/g_\sigma = 0.600$ and $g_\omel/g_\omega = 0.653$. That is, for 1.5$\Lambda$ for instance, the coupling strengths are taken to be $g_\sigl/g_\sigma = 1.5 \times 0.600$ and $g_\omel/g_\omega = 1.5 \times 0.653$. It is found that the neutron star masses are substantially reduced as the $\Lambda$-couplings change from $1.5\Lambda$ to $0.5\Lambda$, roughly corresponding to the uncertainty in $\Lambda$-coupling. Such behavior can be interpreted by the consistency between the EoS and $\Lambda$-coupling strength. With the weakening of $\Lambda$-coupling, which is equivalent to reducing the $\Lambda$ repulsion as well as the Fermi energy, more and more neutrons will be transferred into $\Lambda$-hyperons and the EoS will become softer and softer. From \figref{fig:TOV}(b) one may find that the 2$M_\odot$ constraint is fulfilled reasonably despite the uncertainty of $\Lambda$-coupling. It is also found that with reducing $\Lambda$-coupling the radii of neutron stars decrease to the minimum first and then keep increasing. The CDF calculations also show that the central densities increase to the maximum first and then keep decreasing with the reduction of $\Lambda$-coupling, being consistent with the radius evolutions. {As shown in the plot such turning points in the mass-radius relations are rather close to the original ones (denoted by $\Lambda$), especially for PKA1, which may imply that the coupling strengths $g_\sigl/g_\sigma = 0.600$ and $g_\omel/g_\omega = 0.653$ are reasonable for the $\Lambda$-coupling in stellar matter as well as in finite nuclei \cite{Glendenning:1993}}.

In summary, we studied the general properties of neutron stars with the inclusion of strangeness-bearing $\Lambda$-hyperon, based on the covariant density functional (CDF) theory, specifically the density dependent relativistic Hartree-Fock theory and the relativistic mean field theory with both nonlinear self-coupling of mesons and density dependent meson-nucleon couplings. The inclusion of $\Lambda$-hyperon in neutron star systems shows substantial effects in softening the equation of state for the stellar matter, as well as in reducing the star mass and radius, especially when the contribution of the Fock channel is included. {It is shown that the properties of symmetry energy at normal density are not enough to predict the mass and radius of neutron star.} The systematical investigations on the consistence of the maximum neutron star mass and $\Lambda$-coupling strength also indicate that exotic degrees of freedom is one of the important factors to provide appropriate prediction on the star mass as well as the radius in {consistency} with recent observations.

\begin{acknowledgements}
The author W. H. LONG acknowledges the support from Prof. Hirokazu Tamura and Prof. Osamu
Hashimoto for visiting Tohoku University and finishing part of the work there. This work was also partly supported by the National Natural Science Foundation of China under Grant No. 11075066, and the Fundamental Research Funds for Central Universities under contract No. lzujbky-2010-25, and the Program for New Century Excellent Talents in University, and the Grant-in-Aid for Scientific Research (C) under Contract No. 22540262 and 20540277 from the Japan Society for the Promotion of Science, and the GCOE program G01 and the Grant-in-Aid for Scientific Research on Priority Areas No.17070001 by MEXT, Japan.
\end{acknowledgements}
%

\end{document}